\documentclass{article}

\usepackage{arxiv}

\usepackage[utf8]{inputenc} 
\usepackage[T1]{fontenc}    
\usepackage{hyperref}       
\usepackage{url}            
\usepackage{booktabs}       
\usepackage{amsfonts}       
\usepackage{nicefrac}       
\usepackage{microtype}      
\usepackage{graphicx}
\usepackage{authblk}
\usepackage[version=3]{mhchem} 
\usepackage{siunitx}
\pdfoutput=1
\title{Direct evidence for low-energy electron emission following O LVV Auger transitions at oxide surfaces}

\author[1,*]{Alexander J. Fairchild}
\author[1,*]{Varghese A. Chirayath}
\author[2]{Philip A. Sterne}
\author[1]{Randall W. Gladen}
\author[1]{Ali R. Koymen}
\author[1]{Alex H. Weiss}
\affil[1]{Department of Physics, University of Texas at Arlington, Arlington, TX 76019, United States}
\affil[2]{Lawrence Livermore National Laboratory, Livermore, CA 94550, United States}

\affil[*]{alexander.fairchild@mavs.uta.edu, chirayat@uta.edu}

\begin{document}

\maketitle

\begin{abstract}
 Oxygen, the third most abundant element in the universe, plays a key role in the chemistry of condensed matter and biological systems. Here, we report evidence for a hitherto unexplored Auger transition in oxides, where a valence band electron fills a vacancy in the 2s state of oxygen, transferring sufficient energy to allow electron emission. We used a beam of positrons with kinetic energies of $\sim$1 eV to create O 2s holes via matter-antimatter annihilation. This made possible the elimination of the large secondary electron background that has precluded definitive measurements of the low-energy electrons emitted through this process. Our experiments indicate that low-energy electron emission following the Auger decay of O 2s holes from adsorbed oxygen and oxide surfaces are very efficient. Specifically, our results indicate that the low energy electron emission following the Auger decay of O 2s hole is nearly as efficient as electron emission following the relaxation of O 1s holes in TiO$_2$. This has important implications for the understanding of Auger-stimulated ion desorption, Coulombic decay, photodynamic cancer therapies, and may yield important insights into the radiation-induced reactive sites for corrosion and catalysis.
\end{abstract}


\section{Introduction}
Low-energy electrons are involved in nearly all of the chemical and biological phenomena underlying radiation chemistry playing a central role, for example, in the radiation-induced damage of DNA \cite{Kumar2019} and possibly the origins of life itself \cite{Boyer2016}. Low-energy electrons are most commonly produced as a result of the ionization of core or valence levels by X-ray photons or energetic charged particles. These already ionized atoms or molecules may become further ionized via Auger processes: where correlation effects associated with the filling of vacant electron states (holes) by less tightly bound electrons results in the emission of an electron, the Auger electron. The kinetic energies of these outgoing Auger electrons are characteristic of the electronic levels involved and form the basis of the Auger electron spectroscopies, which have found widespread application in the analysis of surfaces \cite{Ramaker1991}. These low-energy electrons emitted as a result of Auger processes have prompted considerable recent interest in radiobiology due, in part, to their association with enhanced cell lethality \cite{Yokoya2017}.  

Here, we present evidence of a hitherto unexplored O LVV Auger transition at oxygen-exposed Cu and Si surfaces and a TiO$_2$ surface. In order to completely avoid the secondary electron background \textemdash which has previously made definitive identification and quantitative investigation of this Auger transition impossible\cite{Shen1991, Ishida2013, Nakajima2004} \textemdash we have utilized positrons with kinetic energies < 1.5 eV to initiate the Auger process via positron-electron annihilation \cite{Mukherjee2010,Mukherjee2011}. The lack of previous work, and the notable omission of the O LVV Auger transition from standard Auger handbooks \cite{Handbook}, illustrates the difficulty that conventional Auger techniques have in studying this increasingly important Auger electron energy range of 0 to 10 eV. The O LVV Auger emission process is initiated when a surface trapped positron annihilates with an O 2s electron. The resulting hole is subsequently filled by a valence electron, causing a third electron in the valence band to be emitted into the vacuum. This process is schematically represented in Fig. \ref{LVV} for the calculated TiO$_2$(110) density of states \cite{Yang2002}. An added advantage of using positron annihilation to initiate the Auger process is the selectivity to the top-most atomic layer due to the trapping of the positrons at the image potential-induced well on the vacuum side of the sample surface \cite{Mehl1990}. The wave function of the surface trapped positron rapidly decays and has appreciable overlap solely with electrons of the surface terminating atomic species. Thus, the majority of the annihilation-induced holes and the resulting Auger electrons originate almost entirely from the top-most atomic layer. 

\section*{Results}
\begin{figure}[ht]
\includegraphics[width=\linewidth]{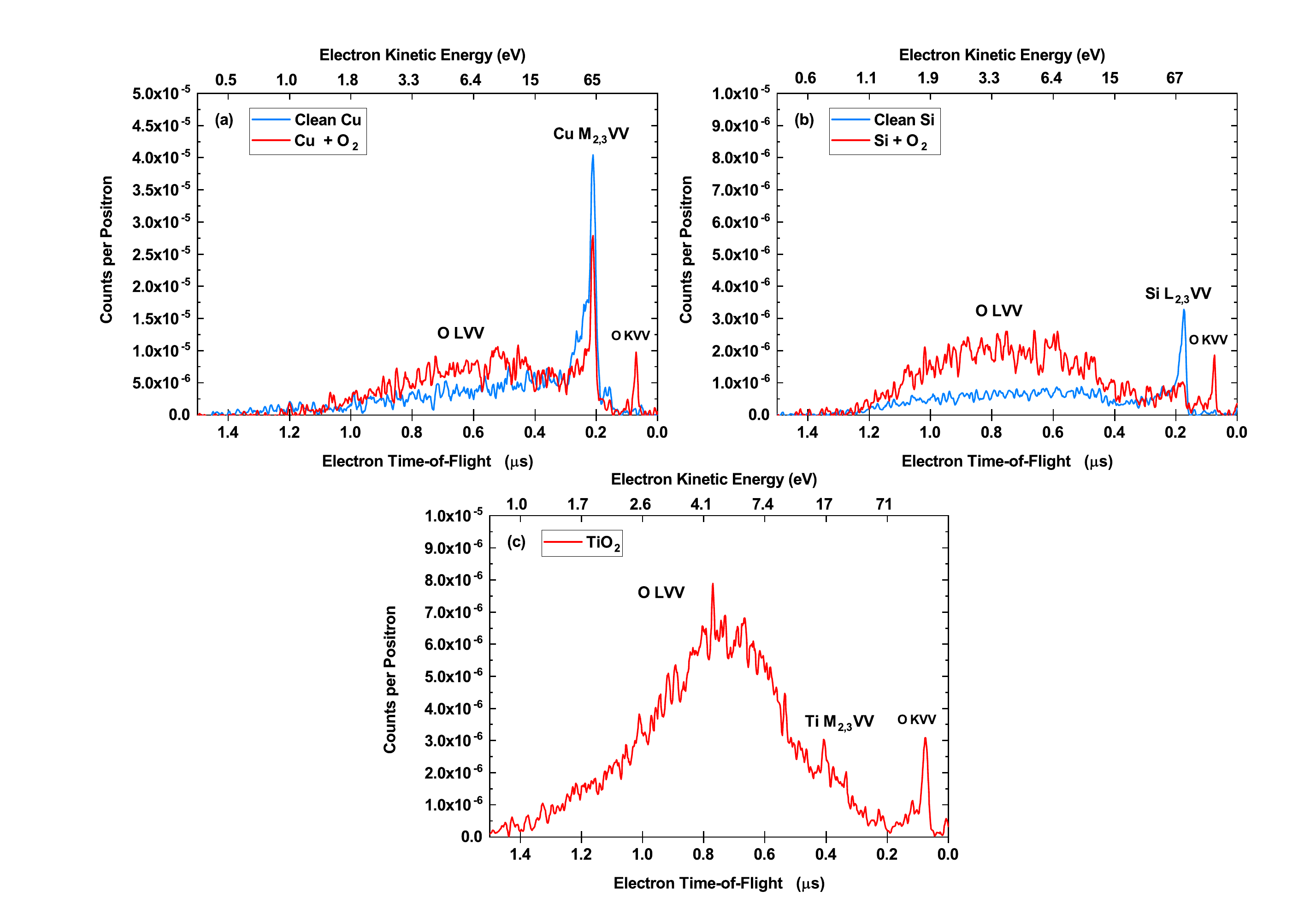}
\caption{Measured ToF-PAES of Cu, Si, and TiO$_2$. ToF spectra of electrons emitted following the Auger decay of positron annihilation-induced holes. The bottom axis is the time the electrons take to travel 1 m. The top axis is the corresponding kinetic energy calculated from the ToFs. Panel (a) compares the clean Cu spectrum (blue) and the oxygen-exposed Cu spectrum (red). The clean Cu spectrum exhibits a Cu M$_{2,3}$VV Auger peak, while the oxygen-exposed Cu spectrum exhibits the additional O LVV and O KVV Auger peaks. Panel (b) compares the clean Si spectrum (blue) and the oxygen-exposed Si spectrum (red). The clean Si spectrum exhibits the the Si L$_{2,3}$VV Auger peak, while the oxygen-exposed Si spectrum exhibits the additional O LVV and O KVV Auger peaks. Panel (c) is the ToF-PAES from TiO$_2$. The spectrum exhibits the Ti M$_{2,3}$VV Auger peak along with the O LVV and O KVV Auger peaks. The low-energy peaks labeled O LVV in each spectrum are primarily due to electron emission resulting from the Auger decay of O 2s levels with some contributions from Auger emission from the metal atoms. This figure has been prepared using OriginPro, Version 2019. https://www.originlab.com/.}
\label{ToF1}
\end{figure}

We have measured the kinetic energies of electrons emitted following O LVV Auger transitions for three surfaces: Cu, Si, and TiO$_2$. Each sample was sputter cleaned and exposed to O$_2$ prior to measurements. Positrons emitted from a $^{22}$Na source were moderated to lower kinetic energies before being transported to the sample surface using a series of electric and magnetic fields. Positrons implanted at low-energies dissipate their initial kinetic energies quickly through inelastic processes and have a high probability of diffusing back to the surface where they become trapped in an image-potential-induced surface state \cite{Schultz1988}. A fraction of these trapped positrons will then annihilate with core electrons creating core holes which may relax via Auger processes. The electron kinetic energies are measured using the time-of-flight technique, where the flight time is taken as the time difference between the detection of the 511 keV annihilation gamma photon and the detection of the outgoing electron. The time-of-flight positron annihilation-induced Auger electron spectra (ToF-PAES) for Cu, Si, and TiO$_2$ are presented in Fig. \ref{ToF1}. The initial measurements of the sputter cleaned Cu (panel (a)) and Si (panel (b)) surfaces exhibit peaks due to the Auger decay of annihilation-induced 3p holes in Cu (Cu M$_{2,3}$VV) and 2p holes in Si (Si L$_{2,3}$VV). The TiO$_2$ spectrum (panel (c)) exhibits peaks due to the Auger decay of annihilation-induced 3p holes in Ti (Ti M$_{2,3}$VV) and 1s holes in oxygen (O KVV). After exposing the Cu (panel (a)) and Si (panel (b)) surfaces to O$_2$ gas, an increase in the integrated PAES intensity below 10 eV is seen alongside the appearance of the O KVV Auger peak. The increased low-energy intensity for oxygen on Cu and Si, and the broad, low-energy peak in TiO$_2$ are each associated with the presence of oxygen on the surface, as indicated by the O KVV Auger peaks. Additionally, the Cu and Si core Auger peak intensities decreased due to the oxygen-induced displacement of the positron wave function away from the Cu and Si atoms of the surface \cite{Fazleev2000,Fazleev2010,Kim1998,Nadesalingam2007b,Nadesalingam2007c}. In the TiO$_2$, the relatively small Ti M$_{2,3}$VV Auger signal relative to the two oxygen-derived Auger signals is consistent with previous findings that the positron wave function overlap for TiO$_2$ is primarily with the surface oxygen atoms \cite{Tachibana2018,Yamashita2017}. An analysis of the integrated PAES intensities shows that the change in the low-energy intensity is 4 times the change in the Cu M$_{2,3}$VV and 18 times the change in the Si L$_{2,3}$VV. This extra intensity cannot be explained by any competing processes, such as inelastic scattering of outgoing Auger electrons \cite{Baglin2000,FIJOL1991}, and is evidence for Auger electron emission following an LVV transition in oxygen adsorbed on Cu and Si surfaces.

\begin{figure}[ht]
\centering
\includegraphics[width=\linewidth]{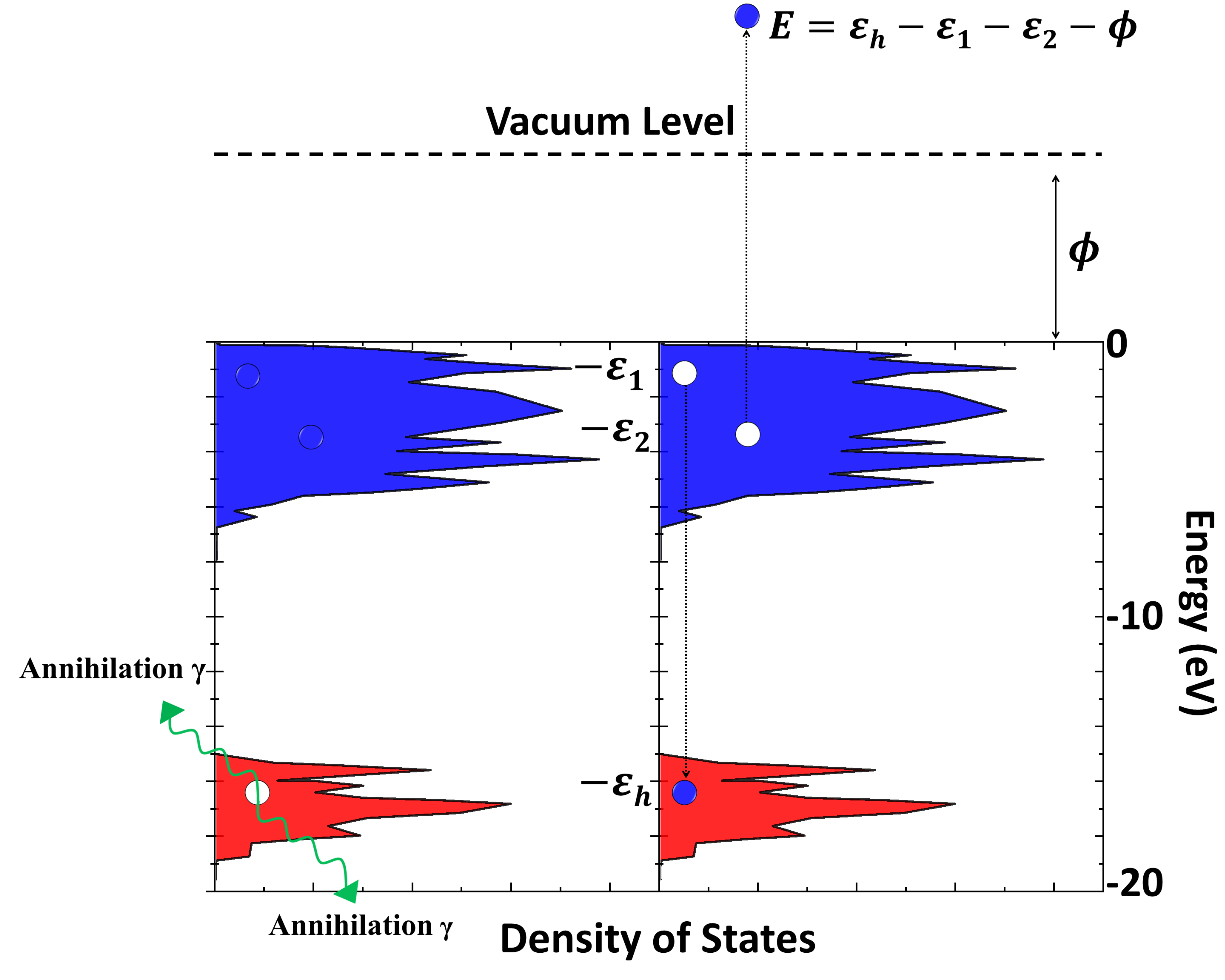}
\caption{Schematic representation of an annihilation-induced O LVV Auger emission process. A surface trapped positron annihilates with an O 2s electron (red) with binding energy $\varepsilon_{h}$ resulting in the emission of two 511 keV annihilation gamma photons. An Auger transition occurs in which a valence band electron (blue), with binding energy $\varepsilon_{1}$, comes to occupy the energy level of the initial core hole. The energy associated with this transition is then coupled to another valence electron in the solid with binding energy $\varepsilon_{2}$, which is emitted into the vacuum. The emitted Auger electron travels with kinetic energy $E = \varepsilon_{h} - \varepsilon_{1} -\varepsilon_{2} - \phi$, where $\phi$ is the energy required to remove an electron from the solid. The density of states, $\rho(\varepsilon)$, shown is for a TiO$_2$(110) surface \cite{Yang2002}. This schematic has been drawn in Microsoft PowerPoint for Office 365.}
\label{LVV}
\end{figure}

\begin{figure}[ht]
\centering
\includegraphics[width=\linewidth]{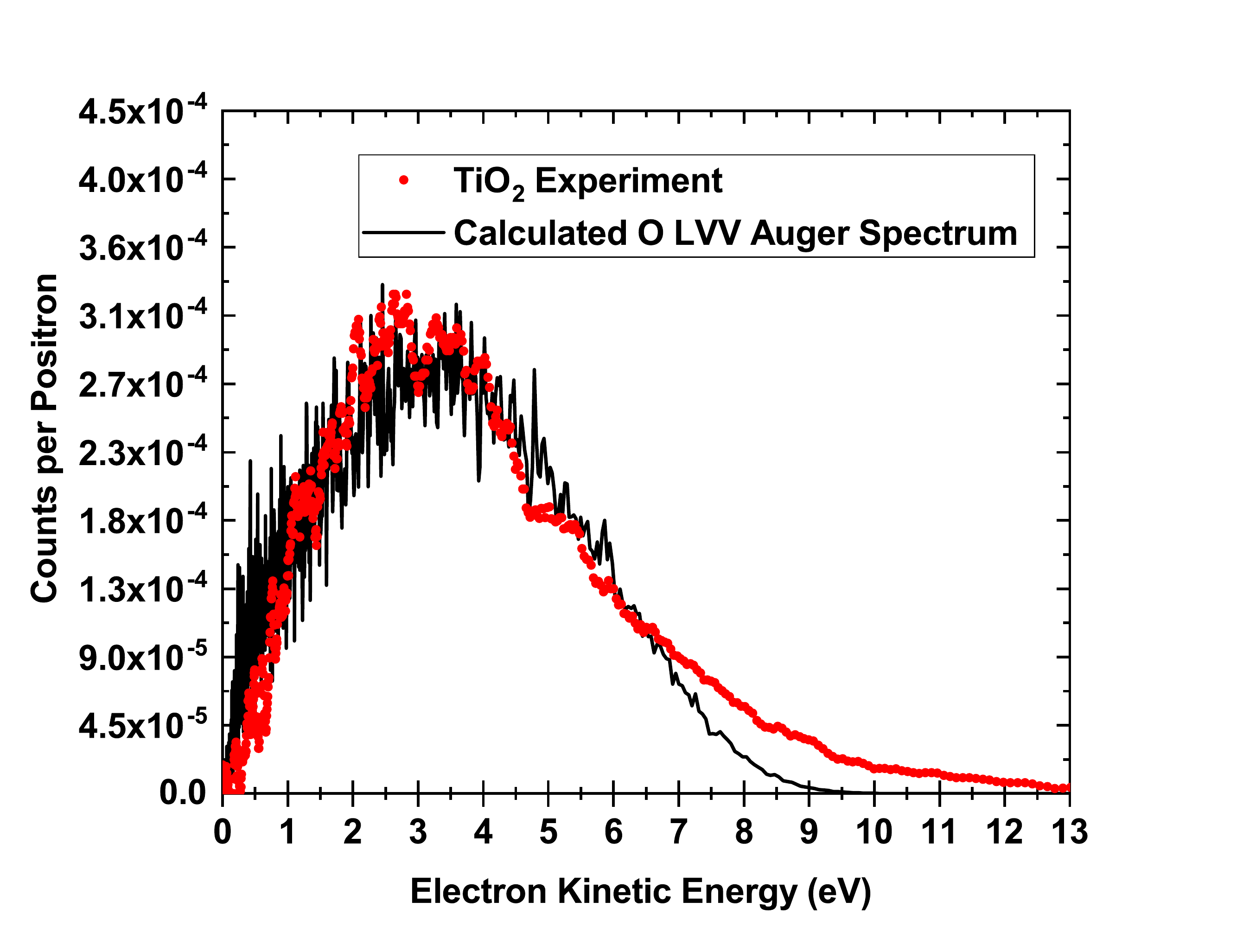}
\caption{Measured and calculated O LVV Auger electron energy spectrum for TiO$_2$. Comparison between the energy converted ToF-PAES spectrum from TiO$_2$ (red) and a calculation of the O LVV Auger spectrum from TiO$_2$ (black) using equation \ref{vvv}. An estimate of the low-energy tail due to inelastically scatted Ti M$_{2,3}$VV Auger electrons has been subtracted from the measured O LVV Auger peak that amounts to $\sim$ 9\% of the total intensity. The calculated spectrum has been broadened using a charged particle trajectory simulation of our ToF-PAES spectrometer. This figure has been prepared using OriginPro, Version 2019. https://www.originlab.com/.}
\label{tio2}
\end{figure}

The energy-converted ToF-PAES spectrum for TiO$_2$, alongside an instrumentally-broadened theoretical calculation of the O LVV Auger electron energy distribution, is shown in Fig. \ref{tio2}. The O LVV Auger electron energy distribution, A(E), was calculated according to: 

\begin{equation}
    A(E) = P_{e}(E)\ \iiint \rho_{h}(\varepsilon_{h}) \rho(\varepsilon_{1}) \rho(\varepsilon_{2}) \delta(\varepsilon_{h}-\varepsilon_{1}-\varepsilon_{2}-\phi-E)\ d\varepsilon_{h}\ d{\varepsilon_{1}}\ d{\varepsilon_{2}}\
\label{vvv}
\end{equation}

where E is the kinetic energy of the emitted Auger electron referenced with respect to the vacuum level, $\varepsilon_h$, $\varepsilon_1$, and $\varepsilon_2$ are the binding energies of the participating electrons and $\phi$ is the energy required to remove an electron from the solid, see figure \ref{LVV}. $P_{e}$(E) is the electron escape probability, which models the probability that an electron has sufficient momentum perpendicular to the surface to escape. The parameters for this empirical function were taken from reference \cite{Chirayath2017}. $\rho_{h}(\varepsilon_{h})$ is the state-dependent density of annihilation-induced holes and $\rho(\varepsilon)$ is the calculated density of states shown in blue in figure \ref{LVV}. $\delta$ is the energy conserving delta function. The density of annihilation-induced holes was approximated using the calculated density of O 2s states, shown in red in figure \ref{LVV}, which corresponds to the assumption of a relatively constant partial annihilation rate. The calculated spectrum was shifted to lower kinetic energies by 8.2 eV to account for the combined effects of the electron work function and final state hole-hole correlation effects. The calculated O LVV Auger spectrum was used as an input to a SIMION\textsuperscript{\textregistered}8.1 simulation of our ToF-PAES spectrometer to account for the effects of instrumental broadening on the outgoing electron energy distribution. Additional details of the simulated ToF-PAES spectrometer and its effects on the calculated Auger spectra can be found in references \cite{Chirayath2017,Fairchild2017}. Finally, an overall scale factor was applied to bring the experimental and calculated peaks into agreement. The excellent agreement between the measured and calculated line shapes provides strong evidence that the observed low-energy peak is a result of O LVV Auger decay processes. We note that the disagreement between 7 and 12 eV between the experiment and the calculation is likely due to final states in which the two holes are in separate oxygen atoms, which is not included in our calculation of the lineshape. These final states have reduced hole-hole repulsion and hence can result in the emission of electrons with higher kinetic energy. This has been identified in other metal oxide systems previously \cite{Salmeron1976} and has been reported in the Auger-like decay of inner-valence holes, which are of predominantly O 2s character, in hydrogen-bonded water clusters \cite{Mucke2010}. Our modelling of the O LVV line shape shows that the maximum kinetic energy available to the outgoing O LVV Auger electrons is $\sim$9.5 eV, when the two final holes are localized at a single atomic site. We have directly measured the maximum energy available following the filling of an O 2s hole to other processes such as Auger-stimulated ion desorption. Our results support the picture put forward by Knotek and Feibelman that energy conservation restrictions are responsible for the weak O$^+$ desorption signal associated with O 2s holes in TiO$_2$ \cite{Knotek1978, Tanaka2000}. 

The ratio of the integrated intensities of the O LVV peak to the O KVV peak is $43 \pm 3$, where the error represents the statistical uncertainty in the measurement. The larger annihilation-induced Auger intensity for the O LVV peak reflects the fact that the ground state positron overlap with the 2s level is higher than the 1s level due to the repulsion of the positron from the positive core. An atomistic calculation shows that the ratio of the number of annihilation-induced O 2s holes to the number of annihilation-induced O 1s holes is 48 (see the Methods section). A comparison of these two ratios indicates that the efficiency of the Auger decay of an O 2s hole is nearly equal to ($\sim$ 90\%) the efficiency of the Auger decay of an O 1s hole. Here, efficiency is defined as the number of electrons detected per initial core hole.

We have performed a detailed calculation of the intensity ratio of the O LVV peak to the O KVV peak taking into account (i) the calculated ratio of O 2s to O 1s annihilation rates, (ii) the probability that an O LVV Auger transition results in an electron with sufficient energy and momentum to escape the material using equation \ref{vvv}, (iii) the effects of the inelastic mean free path of the escaping electron \cite{Chirayath2017}, and (iv) the transport efficiency through our spectrometer \cite{Chirayath2017}. This analysis indicates that the reduced efficiency of the O LVV process relative to the O KVV process is principally due to transitions which do not result in electrons with sufficient energy and momentum normal to the surface to escape. The detailed model yields a ratio of the intensity of the O LVV to the O KVV peaks of 40, which compares favorably with the measured ratio of $43 \pm 3$. We note that in our modelling we have assumed that the Auger decay probability for the L shell vacancy is equal to that of the the K shell vacancy. The agreement between our measured and theoretical ratio supports this assumption. It has previously been shown that the Auger decay probability for the oxygen K shell hole is close to 1 \cite{McGuire1969,Bishop1969}. Hence, we conclude that the Auger decay probability of O 2s holes is also close to unity.

\section*{Discussion}
The unambiguous identification of this previously unexplored, low-energy Auger emission process has implications for photodynamic cancer therapies because O LVV Auger decay (1) is an efficient mechanism for the emission of low-energy, genotoxic electrons and (2) leads to the creation of chemically active, multi-hole final states in localized oxygen atoms. Since TiO$_2$ is widely used in biomedical applications, and low-energy electrons play a crucial role in the nascent stages of DNA radiolysis through dissociative electron attachment (DEA) \cite{Boudaiffa2000}, it is essential that the various mechanisms that can produce low-energy electrons in TiO$_2$ be identified and thoroughly understood. In particular, TiO$_2$ nanoparticles have recently been used in photo-assisted cancer therapies which utilize the emission of low-energy electrons from TiO$_2$ to produce reactive oxygen species \cite{Juzenas2008,Chu2019}. Finally, the final state of the O LVV Auger process is an oxygen atom with two valence band holes. These two-hole final states, which are localized at a single oxygen atom, are believed to enhance cell lethality aiding in the therapeutic potency of photon-induced cancer therapies. These chemically active final states contribute to the enhanced production of cytotoxic free radicals \cite{Saito2014} and to the Coulombic explosion of surrounding water molecules producing further reactive oxygen species \cite{Jahnke2010,Tavernelli2008}.

In conclusion, our work represents the first direct investigations of the emission of low-energy electrons as a result of O LVV Auger transitions in condensed matter systems and may be of significant importance in studies of Auger-stimulated ion desorption and photodynamic cancer therapies. These investigations on oxygen adsorbed surfaces of Cu/Si and TiO$_2$ were made possible by eliminating the large, primary beam-induced secondary electron background by using matter-antimatter annihilation to initiate the Auger process. The success in reproducing both the experimental lineshape of the O LVV Auger peak and the ratio of O LVV to O KVV intensities demonstrates that the picture put forward, that the Auger decay of O 2s holes in TiO$_2$ efficiently results in the emission of low-energy electrons, is correct. Thus, the observation of low-energy electron emission following an O LVV Auger transition from various systems, namely oxygen adsorbed metal(Cu) and semiconductor(Si) surfaces as well as from an oxide surface, shows that the process is ubiquitous; hence, may play an important role in various physical, chemical and biological phenomena initiated by radiation-induced low-energy electron emission.

\section*{Methods}
\subsection*{ToF-PAES and sample preparation}
The experiments presented in this manuscript were carried out using the University of Texas at Arlington's positron beam system. The system is comprised of three parts: a positron beam with magnetic transport, a ToF energy spectrometer, and a sample preparation chamber. A more complete description of the system and its capabilities is provided in reference \cite{Mukherjee2016}. Positrons from a $^{22}$Na source are moderated using a thin tungsten foil in transmission geometry before being magnetically guided to the sample. A permanent samarium–cobalt magnet is mounted a few mm behind the sample which focuses the incoming positron beam onto the sample while also parallelizing along the beam axis the outgoing positron-induced electron momentum. The ToF of the electrons is measured as the time difference between the detection of the 511 keV annihilation gamma rays by a fast scintillator, BaF$_2$ or NaI(Tl), and the detection of the electrons by a microchannel plate (MCP). The sample chamber is kept at a base pressure less than \(10^{-8}\) Pa. The incident positron beam energy was measured to be less than 1.5 eV using a retarding field analyzer. The polycrystalline Cu and Si(100) samples were sputter cleaned every 24 hours before exposure to \num{1.8e3} and \num{2.7e5} Langmuir of O$_{2}$ respectively. A rutile TiO$_2$(110) sample, purchased from Sigma-Aldrich, was sputter cleaned then annealed at 875 Kelvin in an O$_2$ environment of \num{1e-3} Pa for 30 minutes prior to measurements.

\subsection*{Theoretical positron annihilation rate}
The positron annihilation rate, $\lambda$, with a given electronic level $i$ is given by:

\begin{equation}
    \lambda_{i} = \frac{\pi r^2_0c}{e^2}\int \mathrm{d}^3\textbf{r}\ n^+(\textbf{r})n^{-}_i(\textbf{r})\gamma(n_i(\textbf{r}))
\end{equation}

where $r_0$ is the classical electron radius, $n^-$ is the electron charge density, $n^+$ is the positron charge density, and $\gamma$ is the enhancement factor. The enhancement factor takes into account the fact that electrons are attracted to the positively charged positron which increases the positron-electron overlap and hence the positron annihilation rate. A standard self-consistent-field atomic program, which has been utilized previously in simulating the two-detector Doppler-broadening spectra \cite{Sterne2002}, was used to calculate the oxygen electron orbitals. The positron wave function was determined from the positron Schr\"{o}dinger equation using the the calculated electron charge densities with the positron-electron correlation parameterization of Sterne and Kaiser \cite{Sterne1991}:

\begin{equation}
V_{e^-e^+}(r_s) = -1.56(arctan r_s)^{-\frac{1}{2}}+0.1324exp\Big(-\frac{(r_s-4.092)^2}{51.96}\Big)+0.7207
\end{equation}

where r$_s$ is the size of a sphere containing 1 electron given the calculated electron charge density, $n_i^-$, i.e. \(\frac{4\pi}{3}r_s^3n_i^-=1\). Finally, we have used the parameterization of the enhancement factor of Barbiellini et al. \cite{Barbiellini1996}:

\begin{equation}
    \gamma (r_s) = 1+1.23r_s-0.0742r_s^{2}+\frac{1}{6}r_s^3. 
\end{equation}

\bibliographystyle{unsrt}  
\bibliography{Main}  

\section*{Acknowledgements}
This work was supported by Welch Foundation grant No. Y-1968-20180324, NSF grants DMR 1508719 and DMR 1338130, and prepared in part by LLNL under Contract DE-AC52-07NA27344.

\section*{Author contributions statement}
A.J.F. and V.A.C. carried out the measurements. A.J.F. calculated theoretically the O LVV Auger spectra from TiO$_2$. P.A.S calculated theoretically the O 1s and O 2s positron annihilation rates. R.W.G., A.R.K., and A.H.W. assisted in the interpretation of the results and preparation of the manuscript.

\section*{Competing interests}
The authors declare no competing financial interests.

\end{document}